# The Documents and Assets Created During the Video Game Production Process

Katy Daiger
Individual Study Project
University of Texas, School of Information
December 7, 2009

**Table of Contents**

## I. Introduction

### A. Purpose

In 1975, when Atari released the home-consumer version of its arcade hit *Pong*, it was probably hard to imagine how huge the video game industry would be in 2009. Even though it is relatively young, the video game industry has quickly grown from small companies operating out of garages to multi-national companies contributing to the billion dollar industry. Pioneers from the 1970's and 1980's have molded video games into the serious business it is today.

The records that have been accumulating with game companies and game professionals over the years are now of interest to archives, and rightfully so, considering the cultural significance of many of those records. Game developers have embraced this partnership with archives, looking to find a place that will care for the records that they hold so dear.

Professor of Archives and Preservation Management at UCLA Anne Gilliland-Swetland has said that archives are "fundamentally concerned with the organizational and personal processes and contexts through which records and knowledge are created as well as the ways in which records individually and collectively reflect those processes" (2000, p. 2). The problem arises, however, when records cannot be identified during appraisal, and an organization's processes are misunderstood. For an industry like the video game industry, where much of their processes and resulting records are so specific to the industry, it can be hard for an archivist to tell what they are looking at and to understand a record's significance in game production. It is, therefore, in an archivist's best interest to better understand the video game industry and it's many components before making selection and appraisal decisions.

The purpose of this paper is to take that first step in helping archivists understand the video game industry by examining the documents and assets created by game companies. This paper is intended as a survey of the records generated during video game production, and an overview of why and how those records are created. It is not intended to be a statement on archiving best practices, but rather a tool for archivists to use when assessing and processing video game collections. It is an overview of how a video game is made and the paper trail left behind that an archivist might encounter.

### B. Research Techniques and Limitations

Research for this paper was conducted over a single semester through the use of observation and informal interviews. The author spent one day a week for five months examining the production life cycle and the documents and assets generated during that life cycle at a game development company in Austin, Texas. The game company had three concurrent projects all at different stages of production during the observation period. Observations were made regarding the type of documents created, the purpose of those documents, version control for assets, and records retention. The staff at the game company was aware of the research being conducted, and was available for informal interviews when the author had questions.

While this paper provides a useful, in depth look at game production, it is still just an example of how one game company handles documentation. There will be many similarities between the documents and assets created at different game companies, but the amount and type of documentation is highly dependent on the kind of game being created. For example, a small puzzle-based iphone game will need different assets than a large PlayStation 3 adventure game based on a movie property. Documentation is also dependent on what parts of a game a developer is hired to create. The company where the research for this paper was conducted outsourced the audio work for their games to a separate company. Subsequently, audio assets were rather sparse during their production process.

## II. Overview of Game Production

The key players in the video game industry are game developers, publishers, platform companies and IP holders, some of which can be one-in-the-same. The game developer is hired by a publisher to make a game, even if the game concept originally comes from the game developer. While the developer is responsible for creating and implementing an appropriate design aesthetic, many times the developer is given material to work from, especially if there is an IP holder involved. For example, a game developer hired to work on a Harry Potter game would be given design parameters to work within.

The publisher has a huge impact on the documents and assets created during game production. Some publishers take a hands-off approach to their relationships with game developers, only wanting to see documentation when milestones are due. Others ask for highly detailed and granular documentation, wanting to see a game developer's decisions and progress at each step. When there is also an external IP holder involved, they too can request documentation as a way to monitor the handling of their IP.

A video game can be made for a number of game platforms, including the PlayStation 3, Nintendo Wii, Xbox 360, PlayStation Portable, Nintendo DS, Apple iphone and more. Each platform brings with it proprietary technology that a game company must license from the console vendor. In addition, most platforms have specifications for how the game should be built and run. Game developers are usually required to submit a copy of their finished game to the console vendor to check for standards compliance.

Within a game company, employees are divided by disciplines. These divisions include animation, art, programming (or engineering), production, quality assurance, design and audio. Small game productions may combine some disciplines, while larger ones may have more granular divisions. The designers are in charge of creating overall gameplay, look of the game and story. The artists and animators are tasked with implementing the designer's vision and creating the art assets. Programmers focus on the technical aspects of the game, and usually work on creating code for the project. The producer manages a project's schedule and budget, and keeps production on time and under-budget. The quality assurance division is tasked with making sure the game works. They monitor and fix bugs, and check for console standards. As the name would suggest, the audio division creates the music and sounds the play throughout a game. The head of each discipline is considered the "lead."

Game creation generally has four stages of production – concept, preproduction, production and post production. While many documents and assets are created during a specific stage of production, their use can span multiple stages. The concept stage is when a game developer creates a new project and pitches it to a potential publisher. During concept, the game company begins to formulate potential production schedules and budgets to show a publisher what will be required time and budget wise to create the game.

Once a game company is signed with a publisher and a project has been approved for production, the preproduction stage begins. During this stage, the look and feel of the game begins to take shape. Design documents are created and technical road maps are laid out. Prototypes are built to test out potential features and assess possible risks. Preproduction is when plans are finalized, so that game developers can focus on building the game during the production stage.

Production is about creating the actual files that will be used to make a working game. This stage is usually characterized by an upswing in the number of people and the amount of time spent working on the project. Game developers closely monitor problems with the game creation process and the game itself to ensure productivity.

Once the bulk of the production work has been completed, the post production stage begins. This is when all the possible bugs with the game are mitigated. The goal is to get the game ready to be shipped to the publisher and console vendor for final approval. If all goes well, the game will then be mass-produced and sent to video game retailers' shelves.

There are also, of course, administrative and managerial tasks conducted year-round since game development companies are full-time businesses. Many of these tasks play an important role in game production, because of their impact on day-to-day operations.

The documents and assets created during video game production generally fall into one of four categories, although many can belong to more than one category depending on their use. The four categories are official, deliverable, functional and administrative. Official documents describe the relationships between publishers and game developers, and lay out the responsibilities of a game development company. These documents include contracts, amendments and milestone charts. Functional documents help communicate the elements of a game, such as the game design document.

Deliverable documents are those required by a publisher to be turned in at different stages of the game production, such as a playable level. These documents are usually decided upon and laid out in the contract between the developer and publisher. Almost all of the documents and assets listed in the preproduction through post production stages of this paper are deliverables in some form or fashion.

Lastly, administrative refers to documents related to running the company, including executive summaries. For the purposes of this paper, only those administrative documents affecting game production will be mentioned.

## III. Documents and Assets

Below is a list of the documents and assets created during the video game production process. As mentioned above, the list is a representation of how one video game development company creates games, and the records they generate during production. Documents and assets are described in terms of their type and use, and where applicable, examples of content are included. Although the majority of this list deals with the company's active records, there is some mention, where possible, about records management techniques. Each document and asset is listed under the game production stage that best fits its purpose and use, although many records can occur at several points in production.

### A. Year-Round

1. Wiki
An internal wiki is used to facilitate communication within the company. Information on everything from meeting notes to consistency rules to company calendars are posted in the wiki. Using this system allows anyone at the company to edit entries and email links to pertinent posts. While it does not allow for much version control, its functionalities, which include powerful search capabilities, make the Wiki a useful tool for the company. The wiki files are saved when the company's server is backed up, and the company almost never deletes an entry. Wiki pages are saved as records of what has been discussed and decided upon in the past. There is an organization system for the wiki pages, which mostly ensures that records related to different projects will appear together.

2. Perforce
Perforce is less a document and more a piece of software. It is a proprietary, version control software that many game companies use for records management. At first glance, the software looks just like Microsoft Windows Explorer, but it is much more. Assets and documents are "checked in" to the Perforce system after creation, where a server database maintains metadata on the files. When someone wants to work on a file, it must be "checked out," so the software can track its use. Other users are then able to see when and if an asset has been checked out.

When edits have been made to a file, and it is ready to be checked back in, the user is prompted to describe what changes have been made. While Perforce knows that edits have happened, it needs the user to make a specific note of those changes. The file is then placed in a file repository that is on a server separate from user's hard drive. If for some reason an earlier version of a file needs to be accessed, Perforce can revert any asset to any previous version.

The entire Perforce directory and database can be saved, and is often done so at the request of the publisher. The Perforce file is usually one of the final deliverables the game developer gives the Publisher. This is done for legal reasons, but also in case a different company works on a game's sequel.

Author's Note: For the rest of this document, if a version control method is not specified for a document or asset, that indicates that Perforce is used to manage the record.

3. About the Studio / Executive Summary
The game development company finds itself having to give a high level description of the overall structure and philosophies of the company on many occasions. Be it during negotiations with a publisher or an industry conference, it is useful to have a PowerPoint presentation on hand to describe the company. The presentation includes information on the number of full-time employees, the backgrounds of the company leads and management, the company's previously published games, and production viewpoints. It is not unusual for eye-catching graphics to be included. The document is often used in its PowerPoint format, but is also exported as a PDF to facilitate viewing on other computers.

4. Studio Policies
Studio policies are the rules and guidelines created by management and leads for how the company will conduct its operations, including organizational charts and hours employees are expected work at different stages of production. The policies are actually a series of documents, each pertaining to a different aspect of operations. While most of the studio policies are recorded in the wiki, some are saved as Microsoft Word documents and Microsoft Excel spreadsheets. When new policies are added, the document is updated and the old version is discarded. Sometimes, however, notes about certain changes are included in the document, such as "added by John Doe on October 10, 2009" or "changed from 10 hours to 9 hours." If an item needs to be deleted, many times the change is simply crossed out to indicate that a policy that once existed should no longer be used.

5. Production Standards
Similar to studio policies, production standards are rules and guidelines for production, including what should happen in what order, design principles, and how a bug should be described in the bug database. Here again, production standards are actually a series of documents that are mostly included in the wiki. Additions and deletions to the standards are handled the same way as they are for studio policies.

6. Interview Notes
One of the few paper documents that are used at the game development company, interview notes refer to the notes taken by management and leads during job interviews. The documents include interviewee answers to questions, as well as the interviewer's thoughts on the candidate. These records are saved incase of a dispute in the hiring or not-hiring of a candidate.

7. Emails
It seems silly to mention, but emails play a huge role for the company. A large amount of information is communicated via email, including meeting agendas, task updates, questions about standards, and overall work flow issues. Almost everyone at the company saves all their emails pertaining to games currently in production. The emails serve as a reminder of what has been discussed and what tasks are still pending. Microsoft Outlook is their email client of choice, which they setup with numerous folders relating to different aspects of their job. Emails are backed up with the large server back up that happens periodically.

**B. Concept Stage**

1. Pitch Document
This documents is used to "pitch" or sell a game idea to a publisher or other funding source in hopes of being hired to create the game. A pitch can be for a brand new, original game imagined by the development company, or as a result of a publisher's request for proposals for an already planned project. The document is usually created using Microsoft Word, Microsoft PowerPoint or Adobe Photoshop, and is often converted to a PDF for compatibility with other computers.

The lead game designers and artists work on the pitch document, trying to include enough information about their vision without overwhelming the viewer. Information about the overall look and feel of the game, the story and characters, and some gameplay features are usually included in the pitch documents, along with an executive summary of the company. Different versions of the document are generally identified in the file name.

2. Demos
In addition to a pitch document, the development company will often send a playable demo of a game. It is used more as a way to showcase the developer's skills than actually test out game features. While it gives the publisher an idea of what the game will be like should the developer be hired, there is not a lot of back-end technical structure. Most of the work goes into the front end.

The demo is playable on the platform on which the game will eventually be released. So, if the company is making a Nintendo Wii game, the demo will be playable on the Nintendo Wii. Generally speaking, however, the demo is not given to a publisher on a Nintendo Wii optical disk, but as raw Wii files, which can only be executed if the publisher has a Wii development kit. Even with this limitation, this is usually not a problem.

Since this is a playable file, people from most of the disciplines work on the demo. A design scheme is created and implemented by the artists, animators and programmers, and then tested by the quality assurance department. Production does not usually take long, and only a handful of people from each discipline are needed.

3. Cash Flow Schedule
This document is often included with a pitch document. It outlines how many people will be needed to work on a potential project, when they will be needed, and associated personnel costs. Other costs, such as technology licenses and outsourcing payments, are included along with an estimate of when the costs will be incurred.

The executive producer creates this document in Microsoft Excel, but also will consult other leads should questions arise about what resources will be needed for making the game. Columns are used to represent months with total costs at the bottom, while rows represent the employees needed for production. Based on the individual costs, an overall budget is assed, along with potential dates for payment installments. This schedule is not necessarily the schedule that is ultimately used, but it still gives the publisher and company an idea of what it will take to make the proposed game. The version of the file is represented in the file name, as well as the game title, expected date of completion, and estimated budget. Example: GameName_RTM_Jul2009_500K(for distro).xls
(See Appendix A for an example of what the cash flow schedule looks like.)

4. High Level Schedule
The high level schedule differs from the cash flow schedule because it focuses more closely on what each person will do during game production, sometimes down to the individual tasks. Here again, though, the executive producer is in charge of this document and usually builds the schedule in Microsoft Excel. Occasionally, Microsoft Project is used instead, but the file is eventually exported as an Excel spreadsheet. The main purpose of this document is to see how tasks will overlap and to make sure there is enough manpower to get the job done. Since the spreadsheet is laid out with calendar dates in the columns and employees listed in the rows, it is an easy way to visualize each person's workload, and rearrange tasks, if necessary.

5. Developer Evaluation Document
Before a publisher is comfortable paying a game development company, they require the developer to complete the developer evaluation document. This begins life as a Microsoft Word document with numerous questions, ranging from company information and IT structure, to their software engineering processes. The executive producer and other leads then create a second Microsoft Word document with answers to all the questions. These answers allow the publisher to feel more comfortable about the game development company and their ability to actually finish the proposed game project.

6. Contract and Amendments
After a publisher has decided to work with a game development company, a contract must be signed between both parties. As with any business endeavor, the contract lays out the responsibilities of the publisher and developer, and the terms under which both are willing to work together.

This document starts life as a Microsoft Word document created by the publisher and is usually sent electronically to the developer. Management and producers review the document, and then make a paper copy, which can be signed. After the publisher has also signed the contract, the developer saves both the signed paper copy and digital version.

Amendments are often added to contracts, especially when time frames change. Each amendment is approved by both parties and is then integrated into the original contract. Amendments can continue throughout the entire production process.

7. Key Deliverables
The most important section of the contract is the list of deliverables for which the development company will be responsible. Deliverables are associated with different milestones, which are usually evenly spaced throughout the production process. At each milestone, specific documents and assets are required to be completed and turned into the publisher. These deliverables can include everything from the story document to character models, and tend to build on themselves as the creation process continues.

The contract usually includes a master list of all key deliverables included at each milestone, with more granular deliverable lists created later. These lists are either created using Microsoft Word or Excel. The company management and leads work on the deliverables list to make sure that each document or asset required at a milestone can actually be accomplished. As production gets underway, most small changes to deliverables are not made in the contract, but in external deliverable lists. Major deliverable changes are treated as amendments to the contract.

## C. Preproduction Stage

1. Game Design Document (GDD)
One of the most essential documents in the video game production process, the game design document lays out overall style, story and gameplay features. It is written by the lead designers, and is one of the first documents written once a game is green-lit. The GDD is created using Microsoft Word and written in a narrative format. The reader is taken step by step through aspects of the game, resulting in a better understanding of the overall feel of the game. Gameplay features, such as player powers, point collection and reward systems, and how the controls will function, are described in terms of their design and mechanics.

The document often includes references to other games with similar styles or gameplay features. For example, a development company wanting to make an action game set in ancient times might list the game *Assassin's Creed* as an influence. Images and sketches are also included to help explain certain aspects of the game.

When changes need to be made to the GDD, a new version is created. Old information is never written over, rather old versions of the GDD are stored in Perforce incase the developer needs to refer back to a precise incarnation. The document also includes a table on the second page, which lists changes that have been made to the document and who made those changes. Microsoft Word's comment feature is used with the document, especially since the GDD usually goes back and forth many times between the developer and publisher before it is completely agreed upon by both sides.

The game design document can almost be classified as serving a functional, deliverable and official purpose. It is one of many documents created to communicate tasks to the development team, and is almost always required in the first or second milestone. Many publishers also use the GDD in an official capacity, relying on it to determine if production has strayed off course or not.
(See APPENDIX B for an example page from of a game design document.)

2. Additional Design Documents
In addition to a GDD, the game development company usually has other design documents that describe specific design aspects, such as in-game camera angles or player power-ups. These documents are very similar to the larger GDD, but are more granular in scope and sometimes use Microsoft Excel instead of Microsoft Word. They also use a similar system of version control.

The amount of additional design decisions depends on many factors about the project, including the game's genre, intended platform, and whether or not the game is an original property. If the game is small in scope, design ideas can be continually added to the game design document. This is rarely done, though. For one, the GDD could potentially become hundreds of pages long. Also, having multiple design documents allows for multiple people to be working on different design elements at once.

Almost all of these documents end up in the hands of the publisher at some point, usually as requirements for different milestones. This lets the publisher reject elements before the development team gets too far into production.

The publisher or IP owner sometimes provides design documents to be used as guidelines or to convey strict design rules. As mentioned above, when making a Harry Potter game, a developer might receive a design document that explicitly describes the powers that Harry Potter can have. Design documents can also contain lists of unanswered questions about different elements, such as "What color should Harry's hat be?" or "Do we have enough memory for this battle scene?" Once a question is answered, design documents are updated, and the question is moved to a section entitled "answered questions."

3. Story / Narrative Documents
A game's story goes through several steps before it can finally be implemented in the game. It starts with a story document or narrative overview, which lays out what happens in the game and who it happens to. The Microsoft Word document sets up the action and the character motivations. It is written in narrative format and reads like a short story. Dialogue is generally not included at this stage.

Next, the game designer and/or writer creates a list of story stubs, or "beats" of the narrative. These are key points in the story that indicate a need for some gameplay element. For example, a stub might be, "Harry Potter talks with a shopkeeper about the Elder Wand," indicating the need for dialogue and cinematics. A stub might also be "Harry goes to Diagon Alley," indicating a new level within the game. This stub document is created either with Microsoft Word or Excel and allows the developer to start to visualizing gameflow.

Lastly, the stubs are used to create a script. This document contains dialogue for all parts of the game, including during cinematics and gameplay. The script is usually created using Microsoft Word, although movie script programs, like Final Draft, are sometimes used. Voice over is later recorded using this script.

All of these documents go back and forth between the developer, publisher and IP holder for review and discussion. The comments feature of Microsoft Word is used so that issues and concerns are not repeated between the three parties. Starting with the overall story instead of jumping straight to a script allows for changes and additions to be incorporated more easily.

4. Character Roster
As would be expected, the game development company generates a list of characters that appear in the game. The list includes where and when a character appears in the game; what abilities and equipment that character has; and any risks associated with creating the character, such as his power using up too much disk memory. A picture is often included in the list, if concept art has already been created for the character.

This document is created in Microsoft Word by the lead designers and artists, and is often used to help assess what character assets need to be made. Sometimes the character roster is also accompanied by a schedule. This identifies the deliverable that each character is expected by, as well as the order in which the developer will work on the characters. When a schedule is involved, the document will often morph into a Microsoft Excel spreadsheet. (See APPENDIX C for an example from a character roster.)

5. Art Style Guide
Sometimes called the art bible, the style guide is a Microsoft Word document used to communicate to the artists and animators what the game should look like and how to achieve that goal. The style of characters, objects and environments is also often included in this document. Reference images are usually included to help communicate specific style elements.

It is important to note the distinction between this document and design documents. The art style guide is about maintaining an overall look, including lighting and textures, while design documents describe the specific components of characters, objects and environments. When thinking of the Harry Potter game example, the art style guide might tell artists to make all elements associated with the Death Eaters dark and scaly, while design documents would spell out which characters are Death Eaters and that all should wear black robes. There is obvious overlap with these two types of documents, but there is a reason why they are separate.

The art style guide includes instructions of how to use the art software to create the various styles. Everything from how to create texture files in Adobe Photoshop to the number of polygons to use in Autodesk 3DS Max is spelled out in the guide for the artists and animators.

6. Concept Art
Concept art is a 2D visual representation of a game element that allows for the look of the element to be tweaked and perfected before building it in-game. Concept artists create these files based on the game designer's ideas using Adobe Photoshop. Most concept art gets saved as JPEG images, so that anyone can view the files, no matter if they have Photoshop or not.

The style of concept art generally mimics a hand-drawn feel, and looks more like an illustration than a computerized image. Concept art is often drawn from reference photos, which can be just painted over in Photoshop. These documents are often saved and used for marketing and advertising the game.
(See APPENDIX D for an example of concept art.)

7. Reference Files
In many respects, almost anything can be used as a reference. Books can be used for story ideas. Physical objects can be references for objects found in-game. Even sound bytes can generate game design. The most common reference files, however, are reference photos and images used by the artists. Quick time and AVI files are also used to illustrate movements and actions of characters and objects.

Many times publishers will give a game developer references either from previous games or existent IP content. Sometimes these reference are just guidelines, while other times game developers are expected to create carbon copies of the originals. 3D models of characters and environments from previous games can also be used as reference files. Many times the game developer can assimilate the 3D references as a new game assets.
(See APPENDIX E for an example of a reference photo.)

8. World Bible
The world bible has a lot of overlap with the game design document and art style guide, but still serves a unique purpose. The document lays out the rules of consistency for the environments of the game. Not only does it list the worlds and locations in the game, but also the affordance of those locations. Returning to the Harry Potter game example, the world bible for this game would include mention of Hogwarts, London and Gringotts bank, and would list rules such as “wizards can only fly with use of a broom” and “the magic carts at Gringotts should always move fast.” These rules help the artists, animators, level designers and programmers understand how to create game assets and the type of physics to allow in-game.

A world bible is most often seen with games that have a large scope, such as massively multiplayer online games, and those games based on established IPs. If a game is small in scope, much of the world bible can simply be included in the game design document. Here again, the advantage of having these documents separated is avoiding an overly large document file.

9. Level Design Documents
This can be a single document or a series of documents that outline gameplay and flow. Using Microsoft Word, designers layout aspects of what a player will do at any given level and how he will progress through the game. Level design focuses on issues like puzzle solving, boss fights, save points, rewards collections and pathways through an environment. Where as the art style guide and world bible are concerned with how an enemy looks and moves, the level design document describes how and where the player and the enemy will fight.

As with other design documents, this one includes references to other games with similar mechanics and gameplay features, and includes a section for unanswered questions. Depending on the scope of the game, multiple level design documents could be needed to articulate a complex system of levels.

10. Paper Designs
Paper designs go hand in hand with level design documents. These are the hand-drawn overviews of levels made by designers. The designs serve as maps of the environment, illustrating items and areas that a player will encounter.

Traditionally, level designs are drawn on graph paper and then scanned to make a JPEG image of the design. Many times the paper version is discarded in favor of the digital file. More and more these designs are being born digital through the use of Adobe Photoshop and a Wacom graphic tablet. The hardware's stylus pen allows a designer to mimic the feel of a pen or pencil, giving the digital renderings that same hand-drawn look. The designs created in Adobe Photoshop almost always get saved as JPEGs and not Photoshop files. (See APPENDIX F for an example of a paper design.)

11. Gameflow Overview
As the different elements of the game begin to take shape, gameflow charts and walkthroughs are created to make sure everyone at the company is on the same page. These documents show the progression through the game from the viewpoint of a player interacting with the game. Everyone from the producer to quality assurance tester references the gameflow document. While Microsoft Word is sometimes used to create these documents, Microsoft Visio flowchart software is the program of choice.

Starting with the opening sequence, a gameflow chart moves from left to right, documenting the different levels a player will encounter and the order in which they are encountered; player objectives at each level; player powers and abilities at each stage; key narrative beats; and any cinematics that are needed to explain the story. Estimates for how long a player will take to play through a level are included on the chart, as well as a graph that maps the combat, emotional and difficulty level at any given point in the game. Flowcharts are sometimes made for individual levels if the game's size calls for the granularity.

12. UI Flowchart
A UI (or user interface) flowchart focuses on the menu screens seen by the player during gameplay. Created by designers using Microsoft Visio or Excel, the UI flowchart reveals what screens are needed for the game and in what order they should appear, including the save menu and home screen. This document is mostly used by other designers during implementation.

13. Technical Design Document (TDD)
The technical design document is written by the technical director, or lead programmer, as an overview of all technical aspects of game production. It includes details of all software, hardware and game engine components that will be used during production, and what those particular tools will provide to the development team. It also includes a brief description of how code will be written for the game, and pipelines for different assets.

The document is updated at different stages of game production. It often goes back and forth between the publisher and developer as tech specifics become better and better defined. For that reason, there is a chart on the first page of the document that describes changes made to the file and who made the changes. The TDD is often broken out into more granular documents that cover one aspect of the game technology. For example, it is possible to have a document solely devoted to the save game system.

14. Prototype

A prototype is a simplified, yet functioning version of the game that is used to explore a particular question about mechanics, such as "is jumping fun." As apposed to a demo, the front end of the game is fairly plain, but the back end is running serious game mechanics. Some assets are pulled from past prototypes of other games to aid in the production process. Most of the front end of a prototype is in the "grey box" stage, which means that levels are mapped out but contain only grey, simple versions of the objects and environments that will eventually fill the space. Even with this monochromatic interface, mechanics can be easily tested and assessed. Other examples for using a prototype are to test combat systems, vehicle movements and camera angles.

Just like a demo, a prototype is an executable file only playable on the platform on which the game will eventually be released. All disciplines work on building the prototype, including the quality assurance team that, despite the small size of the file, double checks that the prototype runs and works the way the developer had expected.

15. Hiring plan

The hiring plan is written by the producer using Microsoft Word, and describes what disciplines need more hires and at what stage in the production those individuals will be hired. This not only shows the publisher what steps the company is taking to get the game made on time, but it also helps the developer keep on top of hiring new staff. It can be easy to overlook the hiring process when there are so many other development tasks that need to be completed at any given moment. Without those extra staff members, tasks can begin to spiral out of control with not enough manpower to get the job done.

16. Risk Registry

This is a list of potential issues that could arise during production and the possible consequences of those problems. The risk registry is originally created in Microsoft Excel by the producer, but is contributed to by the entire development team. The registry includes assessments of the probability of the risk occurring and ways to mitigate the problem.

There is also a section of the document that lists each attribute to describe a risk, and the possible values for those attributes. Risks are identified as "open" until they are either solved or the issue has passed by without any damage. Since the risk registry is structured to show all changes in risk statues, only one version of the document is created. That version is constantly updated with additions and deletions.

## D. Production Stage

1. Art Asset Tracking Document
Different art assets are submitted to the publisher for approval at various milestones, and they are often returned with comments regarding tweaks that need to be made. The art asset tracking document lists all such assets and when they are due to be submitted, as well as information on assets that have already been submitted. The producer, lead designer and lead artist maintain the list in Microsoft Excel as a way to keep an eye on the numerous assets that go into a game.

The document is usually included with each milestone, and sent back and forth between the developer and publisher as new assets are approved. Since the document is supposed to reflect the various updates and changes that have been made to the list, only one version of the spreadsheet is used and repeatedly updated.

Sometimes there is a need to break out the tracking document into specific art asset, such as character or weapons tracking. These are treated as separate documents and not as different sheets within the same Excel spreadsheet.

2. Models
Before an art asset can be included in a game, it must be created outside of the game engine. Artists use Autodesk 3DS Max to build models of various assets, such as objects or sets. During this stage, a file is saved as a .MAX, representing the 3DS Max project file. Once the model is complete, it is exported as a .S3D file, which is an intermediary, proprietary format used by the developer's game engine. The .S3D file is loaded into the game engine, and immediately converted into one of three model files again specific to the engine.

While most models are saved in highly proprietary formats, they are also sometimes saved as PDFs or .OBJ files. Both of these files can be used to show someone a 2D representation of a model when that person does not have any of the proprietary software.

3. Textures
Models within the game do not automatically have skin. They need to be given textures before they will look like recognizable objects. Artists using Adobe Photoshop create these texture files. They are saved as .TGA files, and imported directly into Autodesk 3DS Max. The .TGA file becomes part of the .MAX project, and is exported to the game engine as part of the .S3D file. Texture files are also sometimes saved as PDFs or .OBJ files if they have to be shown to someone without 3DS Max.
(See APPENDIX G for an example of an art asset at various stages of modeling and texturing.)

4. Game Engine
The game engine is essentially a set of tools for creating a video game. It is a third-party, proprietary piece of software that has already built much of the code required to run the components of the game, including animating models, physics and graphical user interfaces. The engine allows the game developers to focus on the content of the game and not necessarily the back end of the game.

Working within the game engine usually means having several assets open at once, such as a character, a model and a set file. The developer uses a script to determine how these three elements will interact with each other. All the files and associated scripts create the actual levels of the game.

5. Code

Programming code is used to create the structure for the game mechanics. While much of the code is already included in the game engine, there are still situations where code needs to be written for the game to function. When necessary, the Microsoft code compiler and editor Visual Studio is used to write C++ code. These assets are saved as .H and .CPP files. The Nintendo Wii-specific editor CodeWarrior is also used to create code for platform requirements and standards.

6. Script

Script is like code, but tells the game content instructions on what to do and how to do it during gameplay. Script files are written using a proprietary language specific to the developer's game engine. The script files are incorporated directly into the engine, and are included with all other assets when the game is built from the engine. A subset of designers known as scripters create the actual script for the game.

Scripts are often saved as simple text files. This can come in handy when someone who needs to review a script does not actually have the license for the proprietary script software. A localization file is a particular type of script file that lists the written text used throughout the game. When a game is sold in a foreign country, the localization file can be easily accessed and updated to include the text in that country's language. These files are almost always saved as a .TXT file, so that an external translator could access them, should the need arise.

7. Flash Animation

Flash animation is a tool to create user interfaces. The program allows for rapid prototyping of the UI to quickly see what may or may not work in the interface. These assets are saved as .SWF files and imported into the game engine to be used in the final product. A designer is often assigned the task of just creating the user interface.

8. In-game Screen Shots

As one can imagine, there are a so many pieces to video game production, that it can be hard to show a third-party the progress that has been made. It can also be cumbersome to send every single file associated with a level to a publisher wishing to see what the game currently looks like. The game development company gets around this by taking numerous screenshots of the game. These JPEGS can show how the look and feel of the game is developing, as well as specific components that have been created. Quicktime movies are also sometimes used to showcase a particular movement or gameplay feature that would otherwise be incommunicable through a still image, such as sword fighting.
(See APPENDIX H for two examples of in-game screen shots.)

7. Microsoft Project
This software is used by the producer to track specific tasks during game production. Microsoft Project is a project management tool that allows the budget and schedule for a task to be monitored simultaneously. Even though there are so many tasks during the video game production, only those tasks that cost the development company more than $250 are tracked. Examples of possible tasks include the addition of a new character or making sure all the dialogue for the game is recorded.

Generally speaking, each task has its own Microsoft Project file. Depending on the task, some Project files will be converted to Microsoft Excel spreadsheets so that more employees at the game company can read and edit the file.

8. White Board Images
Most meetings at the game company are conducted near a white board. New ideas are talked through with the help of the dry erase board, which is often not erased till days later. When it is time to erase the board in favor of new ideas, digital photos are taken of the white board's contents as a record of what was discussed. The photos are saved as JPEGs and most are uploaded to the wiki with a description of the meeting from which they were generated. The images are backed up when the wiki is backed up. Almost everyone at the company has either created white board content or taken photos of a board and uploaded them to the wiki. The lead running the meeting, though, will often take charge of white board duties.

9. Milestone Feedback
When the developer turns in the documents and assets required for a given milestone, the publisher reviews the files and then sends feedback for the development company. Feedback can be either positive or negative, and often is a mixture of both. It is not uncommon for feedback to include instructions for how to improve different assets. The feedback is generally in the form of a Microsoft Word document, but can also be the original document with comments tracking. Sometimes feedback is communicated via email.

10. Console Vendor Standards
When a developer creates a game for a specific platform, not only will they have to secure a license and development kit, but they will also be required to implement certain standards in the game design. These standards include how the controllers can function and what title screens must appear at the beginning of the game. When a license for the Nintendo Wii is purchased, it includes a collection of PDF documents explaining the various standards required of the company. These PDFs are used by the quality assurance team to make sure that the game complies with the standards.

The last stop a game makes before hitting the retail shelves is the console vendor, who looks over the game to ensure all standards have been fulfilled. Before then, the game developer can communicate directly with the console vendor through email with questions like, "If we did this in the game, would we be disqualified?"

11. Game Build and Build Notes
As different game assets are created and integrated into the game engine, the game becomes more and more complete. Some version of the game is usually required with each milestone, so that the publisher is able to see progress being made.

The build of the game is essentially when all the current assets and components included in the game engine are combined together to run on their own. Once built, the game becomes playable on its intended platform, and functions like a regular console game. The build is sent to the publisher in two forms via FTP site– as .RAR compressed files and .RVM large disk image files. It is up to the publisher which one they want to use.

In addition to the actual game files, there is always a build notes document included with the collection. This Microsoft Word document identifies what game elements are included in the latest build, and the technical requirements for executing the build files. Quality assurance is expected to give the build a last look over to make sure all the correct files are included and that the RVM actually runs.

**E. Post Production Stage**

1. Bug Database
At the end of the production life cycle, the game development company is mostly concerned with identifying and fixing bugs. When a bug is found, it is recorded in the bug database, a web-based database that is proprietary to the publisher, but accessible by all employees at the development company. Bugs can range from a "Class A" bug, which must be fixed, to a "Class C" bug, which the development team would like to fix, but are not as concerned about.

Simultaneous to the quality assurance's efforts to find bugs is the publisher's similar search. The publisher looks for problems and inconsistencies with the game, and communicates these issues to the developer through the bug database. In many cases, a publisher and developer are in different cities, even different countries, so having the publisher input bug information directly into the bug database is an advantage. The developer frequently checks the database for new entries, and assigns a member of the development team the task of assessing and fixing each new bug.

Each bug has its own record with information on the type of bug; where it occurs in the game; who found it; and who is going to fix it. The status of a bug indicates whether it is still unsolved or "open," or if it is solved and "closed." Bug records are not deleted from the database after they are solved to insure that there is a history of what has been found and accomplished.

If need be, the contents of the bug database can be downloaded as a Microsoft Excel spreadsheet. Since access to a publisher's proprietary bug database can be denied after a game is finished, the Excel version is often the only version the game developer can keep.

2. Burn Down Chart
The purpose of the burn down chart is to monitor how many bugs each employee is working on, and to estimate the date when all the bugs will be fixed. The producer uses Microsoft Excel to monitor the bug ceiling, or the number of bugs a person can be responsible for at any given time. If an employee is highlighted red, it means the producer may have to decrease that person's workload. Highlighted yellow means that an employee is nearing the bug ceiling, but is not there yet. The bug ceiling decreases as the game gets closer to completion. There is only one version of the document, which is updated each week with new bug counts.
(See APPENDIX I for a snapshot of the burn down chart.)

3. Focus Group Movies
It is not unusual for the game development company to hire a marketing firm to focus test a game before its release, although many times focus testing is left up to the publisher. Focus testing generally involves a small group of age-appropriate volunteers who play the game and discuss what they did and did not like about it. This information is useful to the developer, who still has a chance to change a few things before the game is finalized. Focus group sessions are taped, and those video files are made available to the developer to watch and review.

4. ESRB Ratings Application
In order to sell a game at most retailers in the United States, the game must be given an Entertainment Software Ratings Board (ESRB) review and official rating. The game developer applies for a review by submitting an application as a Microsoft Word document, along with 30 minutes of footage of typical gameplay. After a few weeks, the ESRB sends the developer another Microsoft Word document that lists the chosen rating and the reasons why. An ESRB ratings application is only submitted when game production has reached the point when no new content is being created.

5. Manual
While most of the layout and final design of a game's user manual is determined by the publisher, the developer is often responsible for the copy. Information about the story, characters and gameplay features are written by the developer, who creates the first version of the draft in Microsoft Word. Almost anyone at the development company can work on the manual, and is often assigned to different individuals who are very familiar with the various aspects of gameplay.

6. GMC Build
The GMC, or gold master candidate, build is the final and complete build of the game that is sent to the console vendor for approval. The GMC build is very similar to previous builds, expect it contains the entire game file. Once the console vendor approves the game for sale, copies are made and distributed by the console company. If for some reason the game is not approved, the developer will have to continue working on the game, which risks pushing their release date.

7. Perforce Archive
Getting the game released to manufacturing, or RTM stage, is part of the game development company's final milestone. This is the point when the console company has approved the game and begun to manufacture copies of it for distribution. Also part of the same milestone is providing the publisher a copy of the game archive. This is the archive of all the files in Perforce, including a copy of the final build of the game and the files required to make that build. It does not include, however, any additional software that might be needed to run the game.

The publisher requests the archive in case there is ever a problem with the game, or if a sequel is made but not by the same developer. The perforce archive also serves as a "receipt" for the production costs the publisher paid the developer.

8. Post Mortem
The post mortem actually happens after a game has shipped. It is a chance for the game development company to reflect on the recent game production, and asses what went right and what went wrong. For those aspects identified as going wrong, possible solutions for the future are outlined.

The post mortem is initially conducted by the leads of each discipline, who collect their thoughts in one Microsoft Word document. A meeting is held with all employees of the company, where the outcome of the post mortem is reported and feedback is solicited. Many of the solutions determined in the post mortem are incorporated into the studio policies and production standards documents, as well as the shared wiki.

## IV. Final Thoughts and Conclusions

The majority of the records listed in this paper are digital files, and more importantly, born digital files. Seeing as how the video game industry is dominated by digital technologies, it would make sense that game developers embrace the world of 1's and 0's when it comes to records management. While there is nothing wrong with how the game community conducts business, its reliance on digital could be a serious hindrance to archives wishing to become the custodians of game records.

Many game assets are created with highly proprietary software, which an archive is most assuredly not going to own. The game industry combats potential technological obsolescence by generating emulators when an old file needs to be accessed. An archive can certainly take this approach as well, but might find itself having to pay a lot of money to find someone to build a game engine emulator.

This may mean that video game archives need to assess what records best represent the "organizational and personal processes and contexts" that Anne Gilliland-Swetland referred to, and focus on obtaining and preserving just those. For example, is it enough to save just the game design documents and the final, consumer version of a video game instead of each milestone build? Maybe not, but those documents would seem to be easier to preserve into the future than .S3D and .SWF files. Maybe the answer is creating PDF and TIFF derivatives of the various assets. An archive is not trying to re-build a game, so is it necessary for them to have every texture file, or are the TIFF images of those textures enough?

Either way, archiving the video game industry seems to call for cooperation between the two communities, and a better understanding of each other's needs. As Gilliland-Swetland would say, "each community [should] learn the others' vocabularies […] and determine what needs to be accommodated and where new practices need to be devised" (2000, p. 1). The video game community can greatly benefit from the care and custodianship of archivists, who will help ensure the history of this important industry is maintained for generations to come.

## V. Work Cited

# APPENDIX A – SNAPSHOT OF CASH FLOW DOCUMENT

**Game Title - Wii**

| | | | | Jul-08 | | Aug-08 | | Sep-08 | |
|---|---|---|---|---|---|---|---|---|---|
| | | | | 1 | 15 | 1 | 15 | 1 | 15 |
| **Platform** | **Group** | **Subgroup** | **Name** | | | | | | |
| Wii | Animation | | John Doe | | 1 | 1 | 1 | 1 | 1 |
| Wii | Art | UI | John Doe | | 1 | 1 | 1 | 1 | 1 |
| Wii | Art | Environment | John Doe | | | | 1 | 1 | 1 |
| Wii | Art | Special Effects | TBD | | | | | | |
| Wii | Code | | John Doe | | 1 | 1 | 1 | 1 | 1 |
| Wii | Code | | John Doe | | 1 | 1 | 1 | 1 | 1 |
| Wii | Production | | TBH - Producer | | 1 | 1 | 1 | 1 | 1 |
| Wii | QA | | John Doe | | | | | | |
| Wii | Design | | John Doe | | 1 | 1 | 1 | 1 | 1 |
| Wii | Design | | Jane Doe | | | | | | 1 |
| Wii | Audio | | Vendor | | | | | | |
| | | | | Kickoff -Jul 15 | | Prepro - Aug 9 | | | |
| | | | Bi-monthly Totals | 0 | 6 | 6 | 7 | 7 | 8 |
| | | | Bi-Monthly Development Costs | $0 | $30,000 | $30,000 | $35,000 | $35,000 | $40,000 |
| | | | **Buyouts** | | | | | | |
| | | | Tech Licenses | | 10,000 | | | | |
| | | | Music Composition | | | | | | |
| | | | Art Vendor | | | | | $20,000 | |
| | | | **Total Buyouts** | $0 | $10,000 | $0 | $0 | $20,000 | $0 |
| | | | **Total Expenses** | $0 | $40,000 | $30,000 | $35,000 | $55,000 | $40,000 |
| | | | **Milestone Payments** | | | $100,000 | | | $100,000 |

# APPENDIX B – EXAMPLE FROM A GAME DESIGN DOCUMENT

## PLAYER CONTROLS

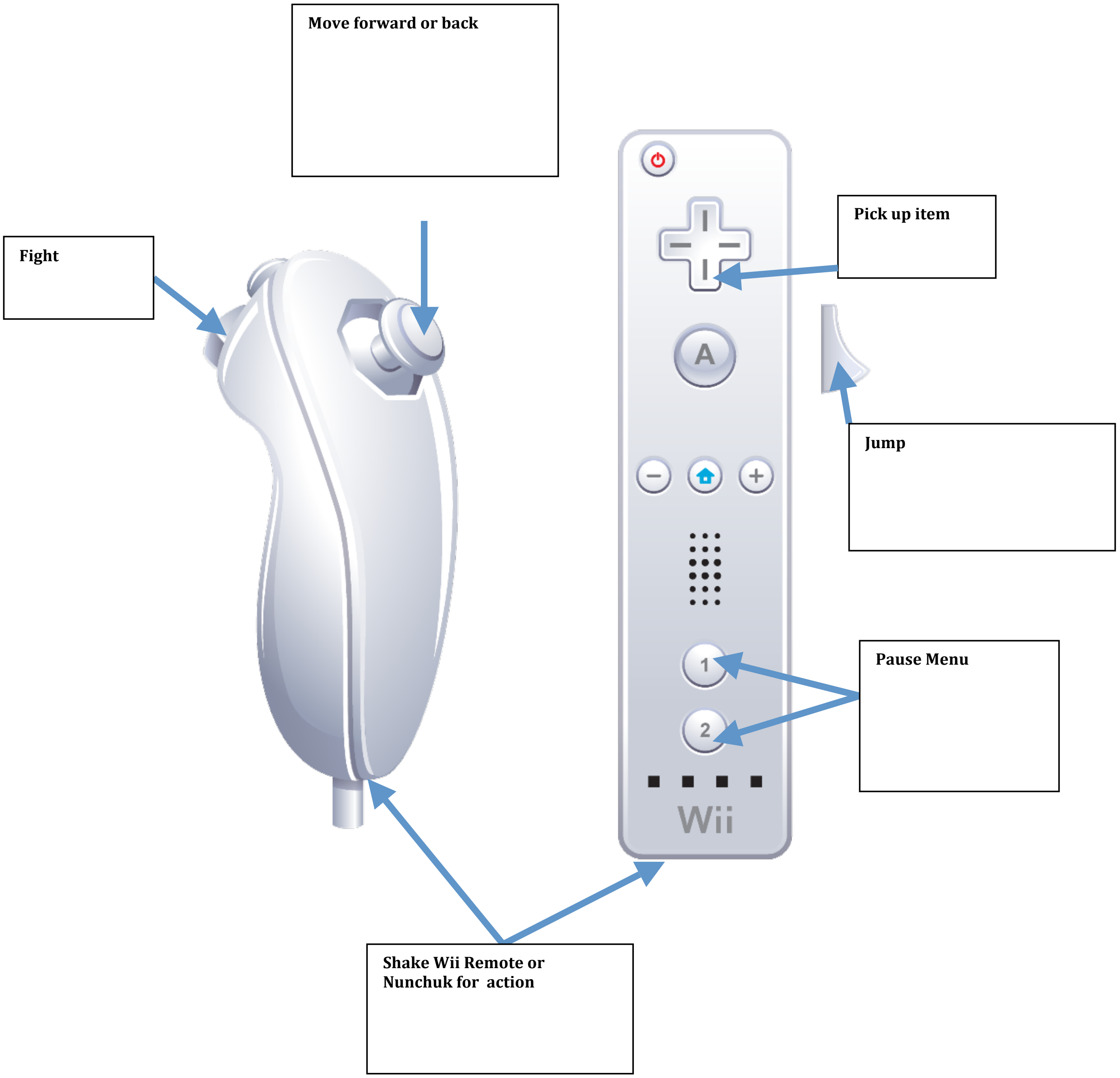

APPENDIX C – EXAMPLE FROM CHARACTER ROSTER

**BOWSER, KING KOOPA**

APPEARS IN

- Level 5

PRIORITY

- High

GAME USAGE

- Enemy for player to fight

ABILITIES AND EQUIPMENT

- Strong fighter
- Mace

OPEN QUESTIONS

- What color should his shell be?

ANSWERED QUESTIONS

- TODO

RISKS AND MITIGATIONS

- Complicated animations
    - Defined early and we should have time to create this character

VARIANTS

- Ghost world version (should be created in black and white)

APPENDIX D – EXAMPLE OF CONCEPT ART

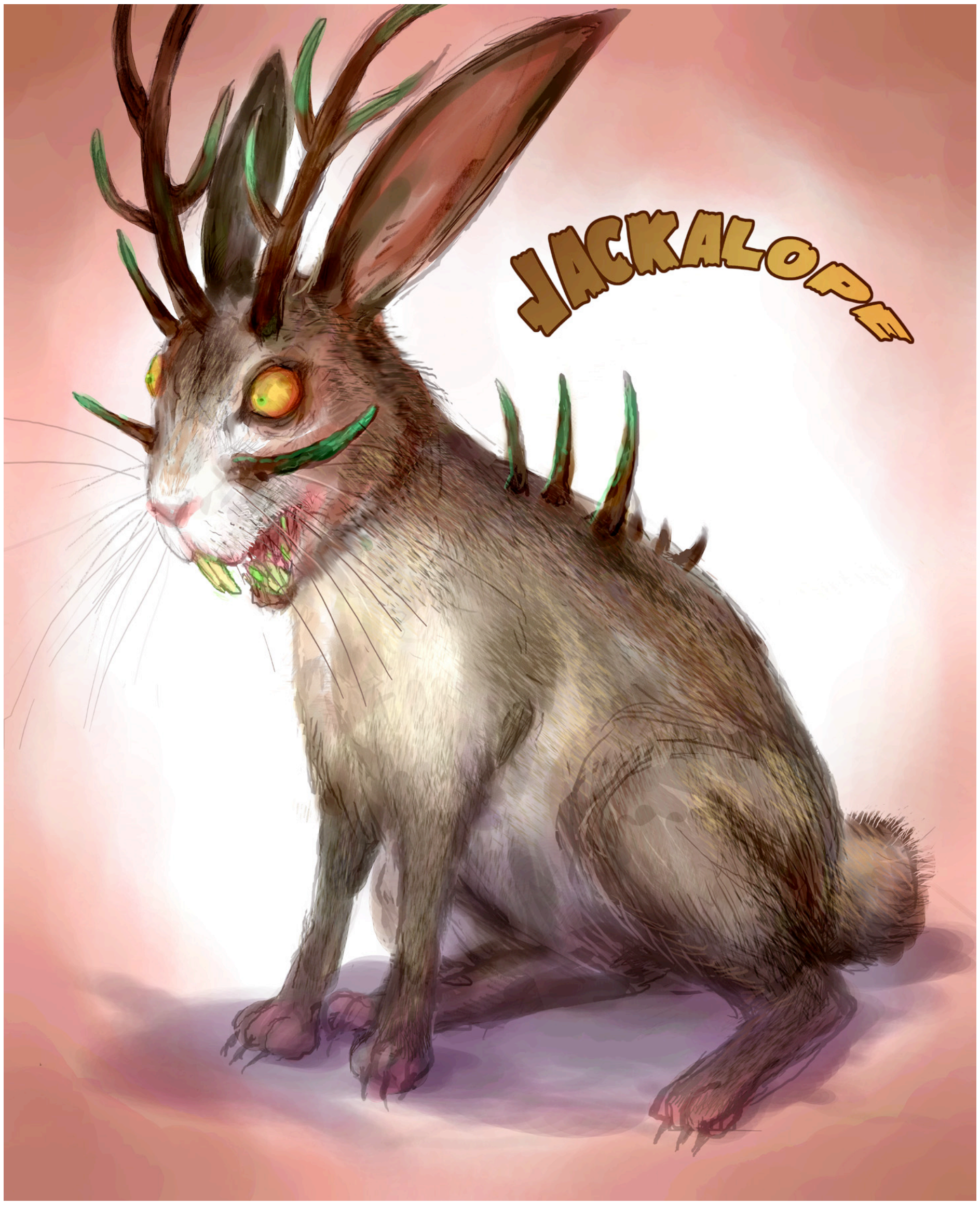

APPENDIX E – REFERENCE PHOTO FOR GAME ENVIRONMENT

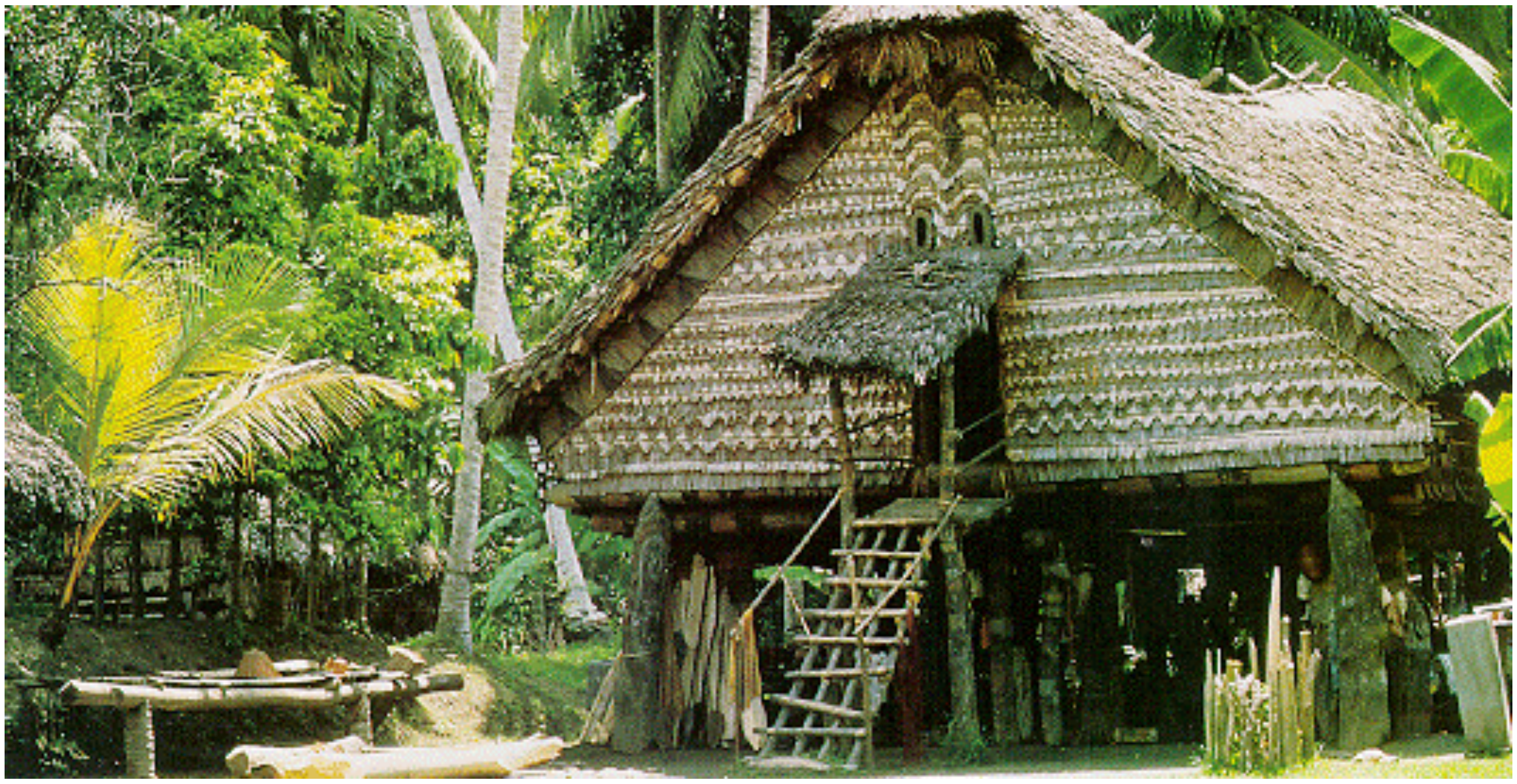

APPENDIX F – PAPER DESIGN EXAMPLE

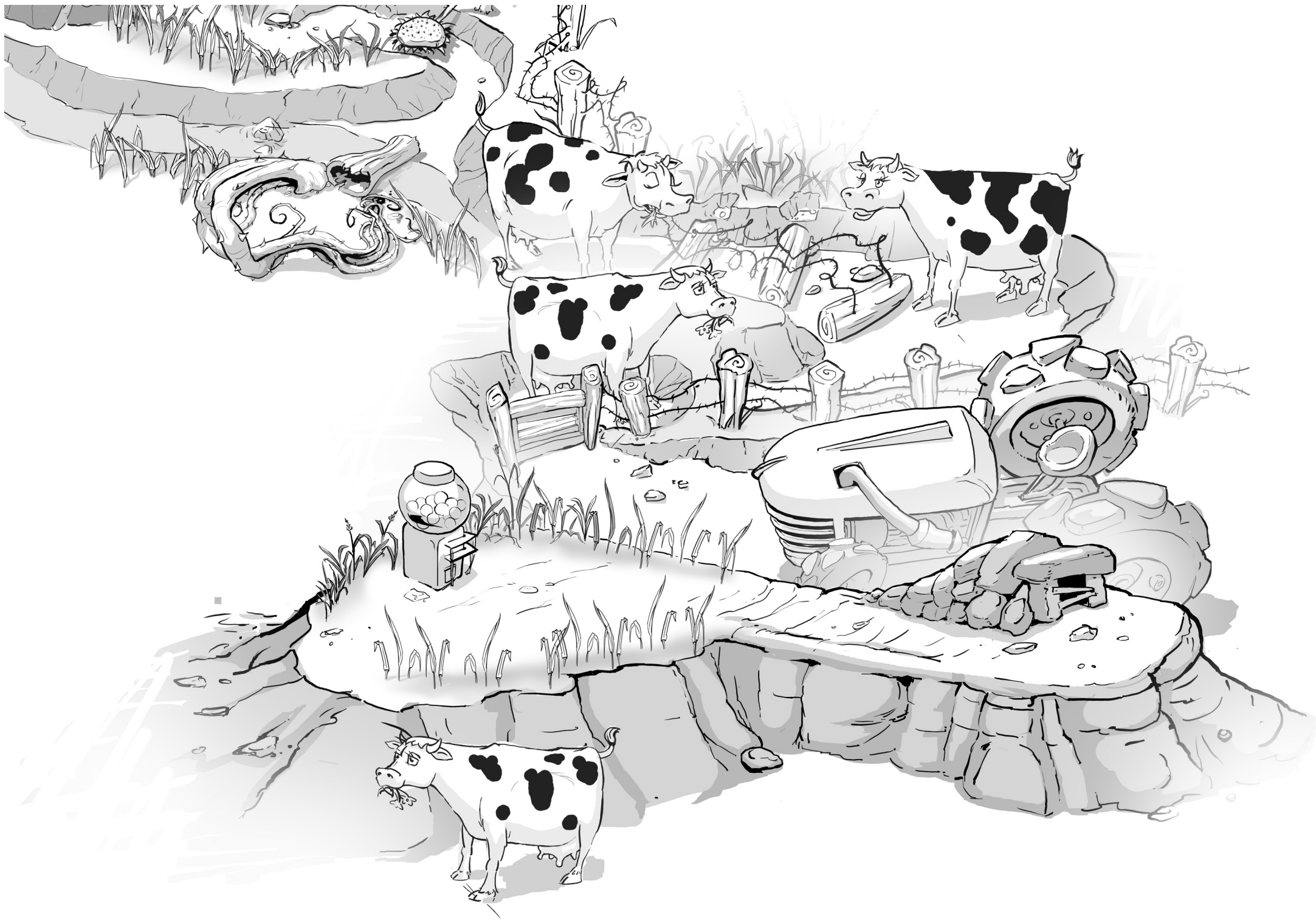

APPENDIX G – EXAMPLE OF ART ASSET AT DIFFERENT STAGES OF MODELING AND TEXTURING

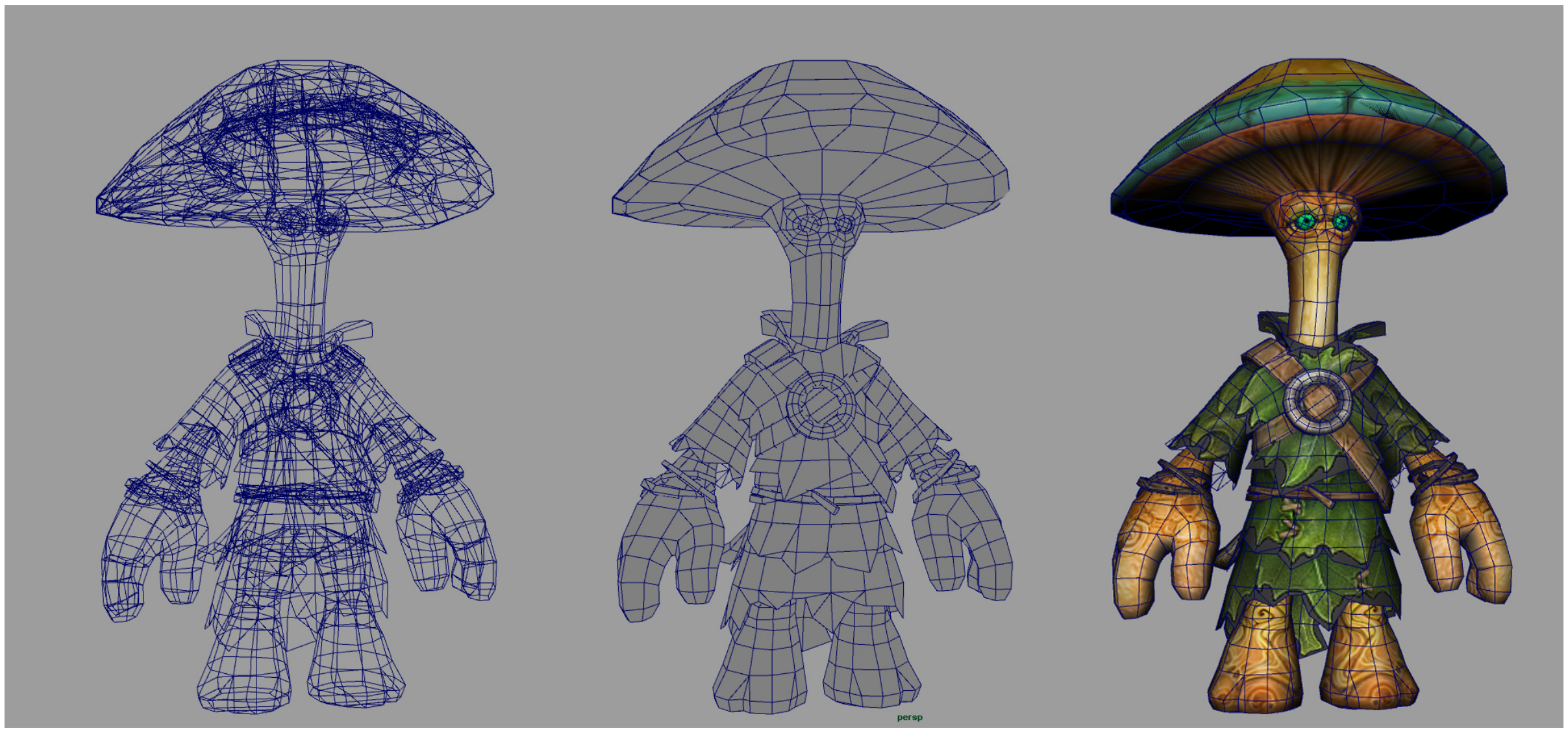

APPENDIX H – EXAMPLE OF IN-GAME SCREEN SHOTS OF A LEVEL AND THE USER INTERFACE

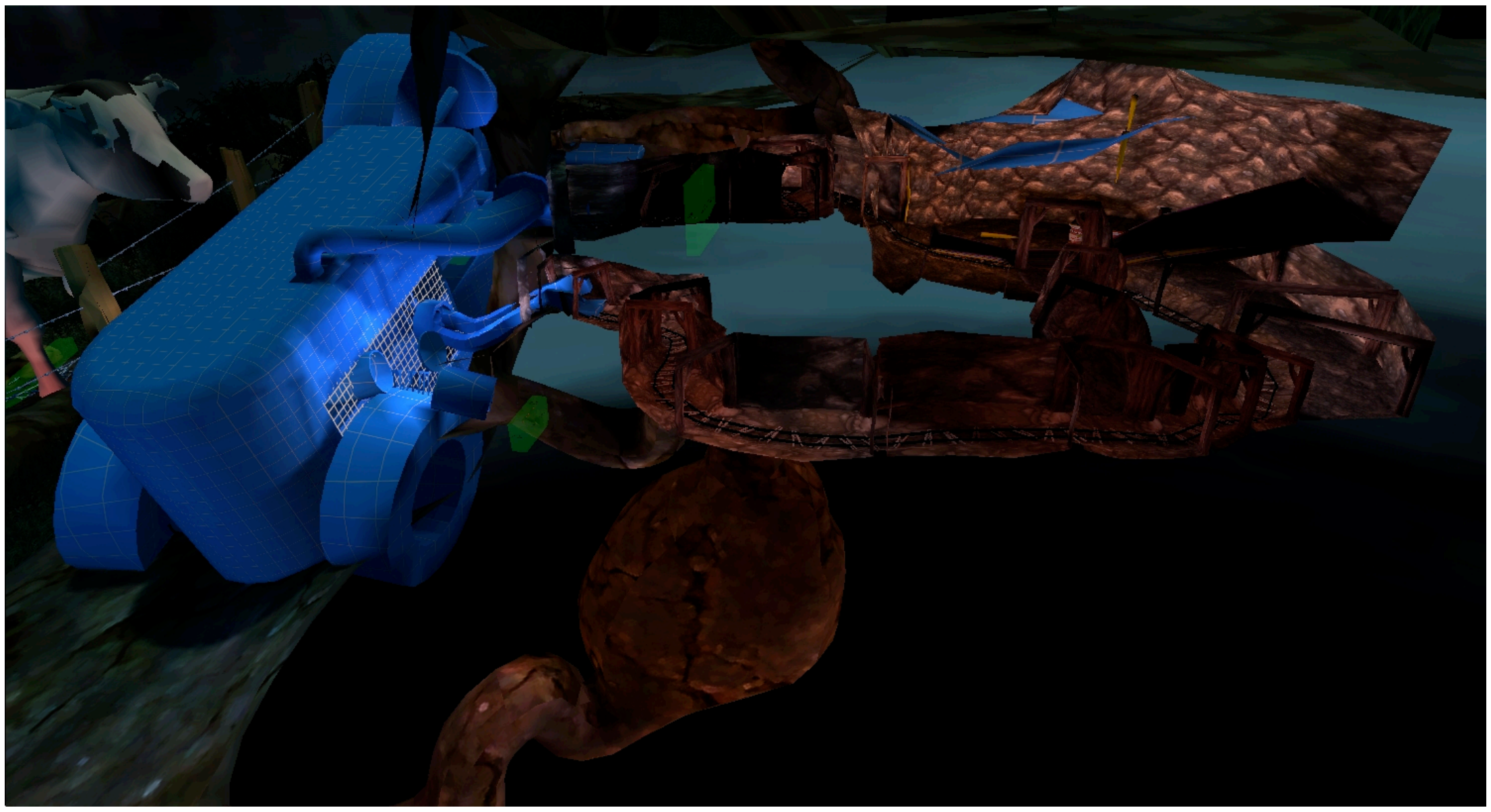

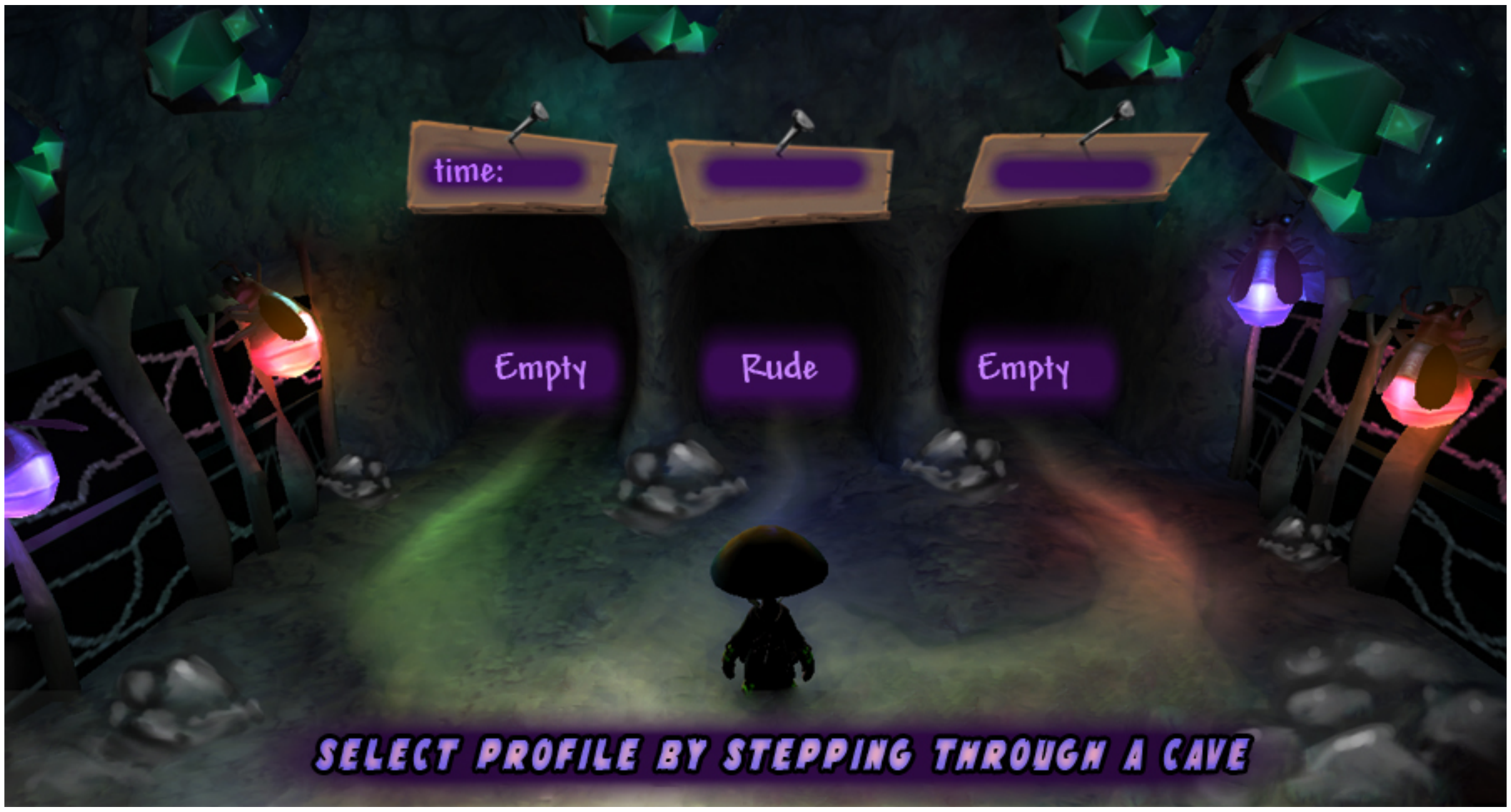

# APPENDIX I – SNAPSHOT OF THE BURN DOWN CHART

| | A | B | C | D | E | F | G | H | I |
|---|---|---|---|---|---|---|---|---|---|
| 1 | Date | 23-Aug-09 | | | | 25-Aug-09 | | | |
| 2 | Bug Ceiling | 0 | | | | 0 | | | |
| 3 | Owner | A | B | Total | Delta | A | B | Total | Delta |
| 4 | Environment Artist | 0 | 0 | 0 | 0 | 0 | 0 | 0 | 0 |
| 5 | Lead Artist | 0 | 0 | 0 | 0 | 0 | 0 | 0 | 0 |
| 6 | Programmer | 0 | 0 | 0 | 0 | 0 | 0 | 0 | 0 |
| 7 | Gameplay Programmer | 0 | 0 | 0 | 0 | 0 | 1 | 1 | 1 |
| 8 | QA Tester | 0 | 0 | 0 | 0 | 0 | 0 | 0 | 0 |
| 9 | Programmer | 0 | 0 | 0 | 0 | 0 | 0 | 0 | 0 |
| 10 | Character Technical Director | 0 | 0 | 0 | 0 | 1 | 0 | 1 | 1 |
| 11 | Creative Lead / Tech Artist | 0 | 0 | 0 | 0 | 0 | 0 | 0 | 0 |
| 12 | Senior Programmer | 0 | 0 | 0 | 0 | 0 | 0 | 0 | 0 |
| 13 | Gameplay Programmer | 0 | 0 | 0 | 0 | 0 | 0 | 0 | 0 |
| 14 | Concept Artist | 0 | 0 | 0 | 0 | 0 | 0 | 1 | 1 |
| 15 | Gameplay Programmer | 0 | 0 | 0 | 0 | 0 | 0 | 0 | 0 |
| 16 | Audio | 0 | 0 | 0 | 0 | 0 | 0 | 0 | 0 |
| 17 | Lead Programmer | 0 | 0 | 0 | 0 | 0 | 0 | 0 | 0 |
| 18 | Gameplay Programmer | 0 | 0 | 0 | 0 | 0 | 0 | 1 | 1 |
| 19 | Gameplay Programmer | 0 | 0 | 0 | 0 | 0 | 0 | 0 | 0 |
| 20 | Tech Director | 0 | 0 | 0 | 0 | 0 | 0 | 0 | 0 |
| 21 | Effects Artist | 0 | 0 | 0 | 0 | 0 | 0 | 0 | 0 |
| 22 | Effects Artist | 0 | 0 | 0 | 0 | 0 | 0 | 0 | 0 |
| 23 | Gameplay Programmer | 0 | 0 | 0 | 0 | 0 | 0 | 1 | 1 |
| 24 | QA Tester | 0 | 0 | 0 | 0 | 0 | 0 | 0 | 0 |
| 25 | Artist | 0 | 0 | 0 | 0 | 0 | 0 | 0 | 0 |
| 26 | Artist | 0 | 0 | 0 | 0 | 0 | 0 | 0 | 0 |
| 27 | Producer | 0 | 0 | 0 | 0 | 0 | 0 | 0 | 0 |
| 28 | Tech Artist | 0 | 0 | 0 | 0 | 0 | 0 | 0 | 0 |
| 29 | Audio | 0 | 0 | 0 | 0 | 0 | 0 | 0 | 0 |
| 30 | Game Designer | 0 | 0 | 0 | 0 | 0 | 0 | 3 | 3 |
| 31 | Artist | 0 | 0 | 0 | 0 | 0 | 0 | 0 | 0 |
| 32 | UI Artist | 0 | 0 | 0 | 0 | 0 | 1 | 2 | 2 |
| 33 | Animator | 0 | 0 | 0 | 0 | 0 | 0 | 0 | 0 |
| 34 | Environment Artist | 0 | 0 | 0 | 0 | 0 | 0 | 0 | 0 |
| 35 | unassigned | 0 | 0 | 0 | 0 | 0 | 0 | 0 | 0 |
| 36 | (blank) | 0 | 0 | 0 | 0 | 0 | 0 | 0 | 0 |
| 37 | Dev Total | 0 | 0 | 0 | 0 | 1 | 2 | 10 | 10 |
| 38 | FIXED QUEUE | 8 | | | | 10 | | | |
| 39 | Total All Bugs | 3089 | | | | 3003 | | | |
| 40 | New Bugs | 86 | | | | 72 | | | |
| 41 | Fix Rate | 96 | | | | 68 | | | |
| 42 | Zero Bug Day | N/A | | | | N/A | | | |
| 43 | Date | 23-Aug-09 | | | | 25-Aug-09 | | | |
| 44 | Bug Ceiling | 0 | | | | 0 | | | |